\documentclass[fleqn,12pt,twoside]{article}
\usepackage{espcrc1}
\usepackage{epsfig}

\newcommand{\gsi}     {$\rm^{a}$}
\newcommand{\wro}     {$\rm^{b}$}
\newcommand{\hei}     {$\rm^{c}$}
\newcommand{\ud}{\mathrm{d}}

\newcommand{\AmS}{{\protect\the\textfont2
  A\kern-.1667em\lower.5ex\hbox{M}\kern-.125emS}}

\title{Statistical hadronization of charm at SPS, RHIC and LHC}

\author{A.~Andronic\gsi, P.~Braun-Munzinger\gsi, K.~Redlich\wro, J.~Stachel\hei}

\begin{document}

\maketitle

\gsi~GSI Darmstadt, Germany,
\wro~University of Wroclaw, Poland,
\hei~University of Heidelberg, Germany \\

\begin{abstract}
We study the production of charmonia and charmed hadrons 
for nucleus-nucleus collisions at SPS, RHIC, and LHC energies
within the framework  of the statistical hadronization model.
Results from this model are compared to the observed centrality
dependence of  $J/\psi$ production at SPS energy. We further provide
predictions for the centrality dependence of the production of open 
and hidden charm mesons at RHIC and LHC.
\end{abstract}

\section{Introduction }

The original idea of statistical hadronization of charm quarks
\cite{pbm1,pbm2} has sparked an intense activity in this field
\cite{gor,gra}.  Initial interest focussed on the available SPS
data on $J/\psi$ production. As we show below, these data can be well
described, but only assuming an enhanced charm cross section compared to
pQCD calculations.
However, the largest differences between
results from the statistical coalescence scenario (or a similar model 
\cite{the}) and more conventional models are expected at collider energies.  
For example, in the Satz-Matsui approach \cite{satz}, one would expect 
very strong suppression compared to direct production of $J/\psi$ mesons 
(up to a factor of 20 \cite{vogt99}) for central Au-Au collision at RHIC
energy. In the present approach this suppression is overcome by 
statistical recombination of $J/\psi$ mesons from different $c \bar c$
pairs, so that much larger yields are expected. We therefore focus in
this note on predictions for open and hidden charm mesons at RHIC and
LHC energy, with emphasis on the centrality dependence of rapidity densities.

\section{Model description} 

The model assumes that all charm 
quarks are produced in primary hard collisions and equilibrate in quark-gluon 
plasma (QGP); in particular, no $J/\psi$ is preformed in QGP (complete 
screening).
All charmed hadrons (open and hidden) are formed at freeze-out 
(at SPS and beyond, freeze-out is at the phase boundary \cite{pbm3})
according to the statistical laws.
Taking into account the measured dependence of the ratio $\psi'/\psi$ on 
centrality, the model is valid for not too small number of participants,
$N_{part}$.
The total number of thermally produced open charm hadrons, $N_{oc}^{th}$ is 
related to the number of directly produced $c\bar{c}$ pairs, 
$N_{c\bar{c}}^{dir}$ as (neglecting charmonia):
$N_{c\bar{c}}^{dir}=\frac{1}{2}g_c N_{oc}^{th}
{I_1(g_cN_{oc}^{th})}/{I_0(g_cN_{oc}^{th})}$, 
from which the charm enhancement factor $g_c$ is extracted. 
$I_n$ are the modified Bessel functions.
The yield of a given species $X$
is then $N_X=g_cN_X^{th}{I_1(g_cN_{X}^{th})}/{I_0(g_cN_{X}^{th})}$
for open charm mesons and hyperons and $N_X=g_c^2N_X^{th}$ for charmonia
 (see ref.~\cite{pbm1,pbm2} for more details).
The inputs for the above procedure are: 
i) the total charged particles yields (or rapidity densities), which are taken
from experiments at SPS \cite{na49,na50a} and RHIC \cite{pho} and extrapolated
for LHC; 
and ii) $N_{c\bar{c}}^{dir}$, which is taken from next-to-leading order (NLO)
perturbative QCD (pQCD) calculations for pp \cite{vogt} (the yield from MRST 
HO parton distributions was used here) and scaled to AA via the nuclear 
overlap function. 
A constant temperature of 170 MeV and the baryonic chemical potential $\mu_b$ 
according to parametrization $\mu_b=1270/(1+\sqrt{s_{NN}}/4.3)$
\cite{pbm4} have been used for the calculations.

\section{Results}

\begin{figure}[htb]
\begin{tabular}{cc}
\begin{minipage}{.6\textwidth}
\centering\includegraphics[width=1.02\textwidth,height=.97\textwidth]{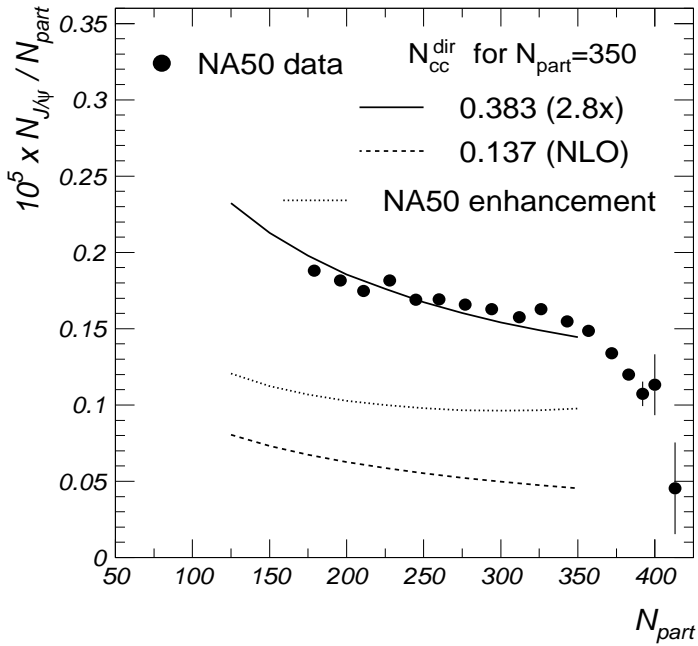}
\end{minipage}  
& \begin{minipage}{.35\textwidth}
\vspace{-11mm}
\caption{The centrality dependence of $J/\psi$ production at SPS.
Model predictions are compared to NA50 data (4$\pi$-integrated).
Two curves for the model correspond to values of $N_{c\bar{c}}^{dir}$
from NLO calculations \cite{vogt} (0.137, leading to $g_c$=0.78) and scaled 
up by a factor of 2.8. The dotted curve is obtained when considering a
charm enhancement as suggested by NA50 \cite{na50b}.
For central collisions ($N_{part}$=350) we use in the calculations 
V=3070~fm$^3$ and  $N_{oc}^{th}$=0.98.}
\label{aa:fig1}
\end{minipage}
\end{tabular}
\end{figure}

We first compare predictions of the model to 4$\pi$-integrated
$J/\psi$ data at the SPS (NA50 data \cite{na50} replotted as outlined in
\cite{pbm2}).  The differences between this calculation and that of
\cite{pbm2} are due to the inclusion in the present version of the
complete set of charmed mesons (open and hidden) and baryons as well
as due to updated values of the inputs (volume and open charm cross
section) as outlined above. In Fig.~\ref{aa:fig1} we present the
comparison for two values of $N_{c\bar{c}}^{dir}$: from NLO
calculations \cite{vogt} and scaled up by a factor of 2.8. Note
that the observed centrality dependence is well reproduced using the
NLO cross sections for charm production scaled by the nuclear overlap 
function. 
To explain the overall magnitude of the data a $N_{c\bar{c}}^{dir}$ increase 
by a factor of 2.8 compared to NLO calculations is needed.

We mention in this context that the observed enhancement of the dilepton 
yield at intermediate masses \cite{na50b} has been interpreted 
as possible indication for an anomalous increase in the charm cross section. 
If, in our calculations, we scale-up $N_{c\bar{c}}^{dir}$ from NLO according
to the NA50 enhancement ("Data/Expected sources") as a function of $N_{part}$ 
\cite{na50b}, the $J/\psi$ data are underpredicted and the $N_{part}$ 
dependence is flatter (see Fig.~\ref{aa:fig1}). 
However, the NA50 suggestion of charm enhancement is controversial, and other 
explanations exist of the observed enhancement
in terms of thermal radiation \cite{rapp,gall}.

\vspace{-4mm}

\vspace{-8mm}
We turn now to predictions for collider energies. For comparison we
include in this study also results at
SPS energy.  The input parameters for these calculations for central 
collisions ($N_{part}$=350) are presented in Table~\ref{aa:tab1}. 
Notice that from now on we focus on rapidity densities, which are the 
relevant observables at the colliders. The results  are compiled in
Table~\ref{aa:tab2} for a selection of hadrons with open and hidden
charm. All predicted yields increase strongly with  energy, reflecting
the increasing charm cross section and the concomitant importance of
statistical recombination. Also ratios of open charm hadrons evolve
with increasing energy, reflecting the corresponding decrease in charm
chemical potential.

\vspace{-4mm}
\begin{table}[htb]
\begin{minipage}[t]{0.43\textwidth}
\caption{Input parameters for model calculations at
SPS, RHIC and LHC.}
\label{aa:tab1}
\end{minipage} 
\begin{minipage}[t]{0.03\textwidth} ~~
\end{minipage} 
\begin{minipage}[t]{0.53\textwidth}
\caption{Results of model calculations at
SPS, RHIC and LHC for $N_{part}$=350.} 
\label{aa:tab2}
\end{minipage} 

\begin{tabular}{|c|ccc|c|c|ccc|}
\cline{1-4} \cline{6-9}
$\sqrt{s_{NN}}$ (GeV) &  17.3  &  200 &  5500 ~  & ~ &
$\sqrt{s_{NN}}$ (GeV) &  17.3  &  200 &  5500  
\\ \cline{1-4} \cline{6-9}
$T$ (MeV)        & 170   & 170   & 170  & ~ &
$\ud N_{\mathrm{D}^+}/\ud y$                  & 0.010 & 0.404 & 3.56 
\\
$\mu_b$ (MeV)    & 253   & 27    & 1     & ~ &
$\ud N_{\mathrm{D}^-}/\ud y$                  & 0.016 & 0.420 & 3.53
\\ \cline{1-4} 
$\ud N_{ch}/\ud y$  &  430   &  730    &  2000  & ~ &
$\ud N_{\mathrm{D}^0}/\ud y$                  & 0.022 & 0.89 & 7.80
\\
$V_{\Delta y=1}$ (fm$^{3}$) &  861   &  1663   &  4564  & ~ &
$\ud N_{\Lambda_c}/\ud y$                     & 0.014 & 0.153 & 1.16
\\ \cline{1-4} 
$\ud N_{c\bar{c}}^{dir}/\ud y$  & 0.064 & 1.92  & 16.8  & ~ &
$\ud N_{J/\psi}/\ud y$  &  2.55$\cdot$10$^{-4}$ & 0.011 & 0.226
\\
$g_c$                           & 1.86  & 8.33  & 23.2  & ~ &
$\ud N_{\psi'}/\ud y$ & 0.95$\cdot$10$^{-5}$ & 3.97$\cdot$10$^{-4}$ & 8.46$\cdot$10$^{-3}$
\\ \cline{1-4} \cline{6-9}
\end{tabular}
\end{table}

\vspace{-12mm}
\begin{figure}[hbt]
\begin{tabular}{lr} \begin{minipage}[t]{0.48\textwidth}
\centering\includegraphics[width=1.06\textwidth,height=1.09\textwidth]{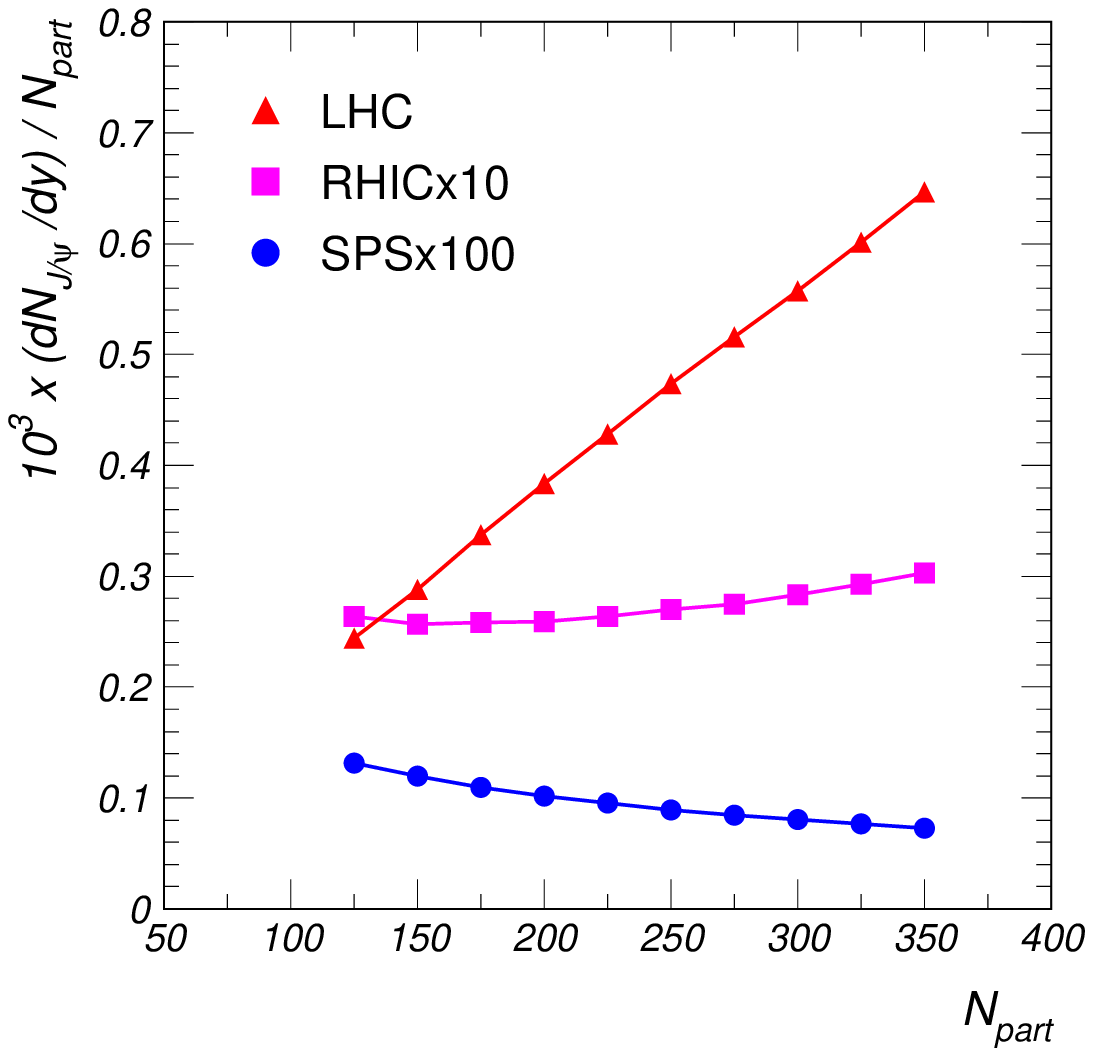}
\vspace{-10mm}
\caption{Centrality dependence of $J/\psi$ rapidity density at SPS, RHIC 
and LHC.} \label{aa:fig2}
\end{minipage} & \begin{minipage}[t]{0.48\textwidth}
\centering\includegraphics[width=1.04\textwidth,height=1.09\textwidth]{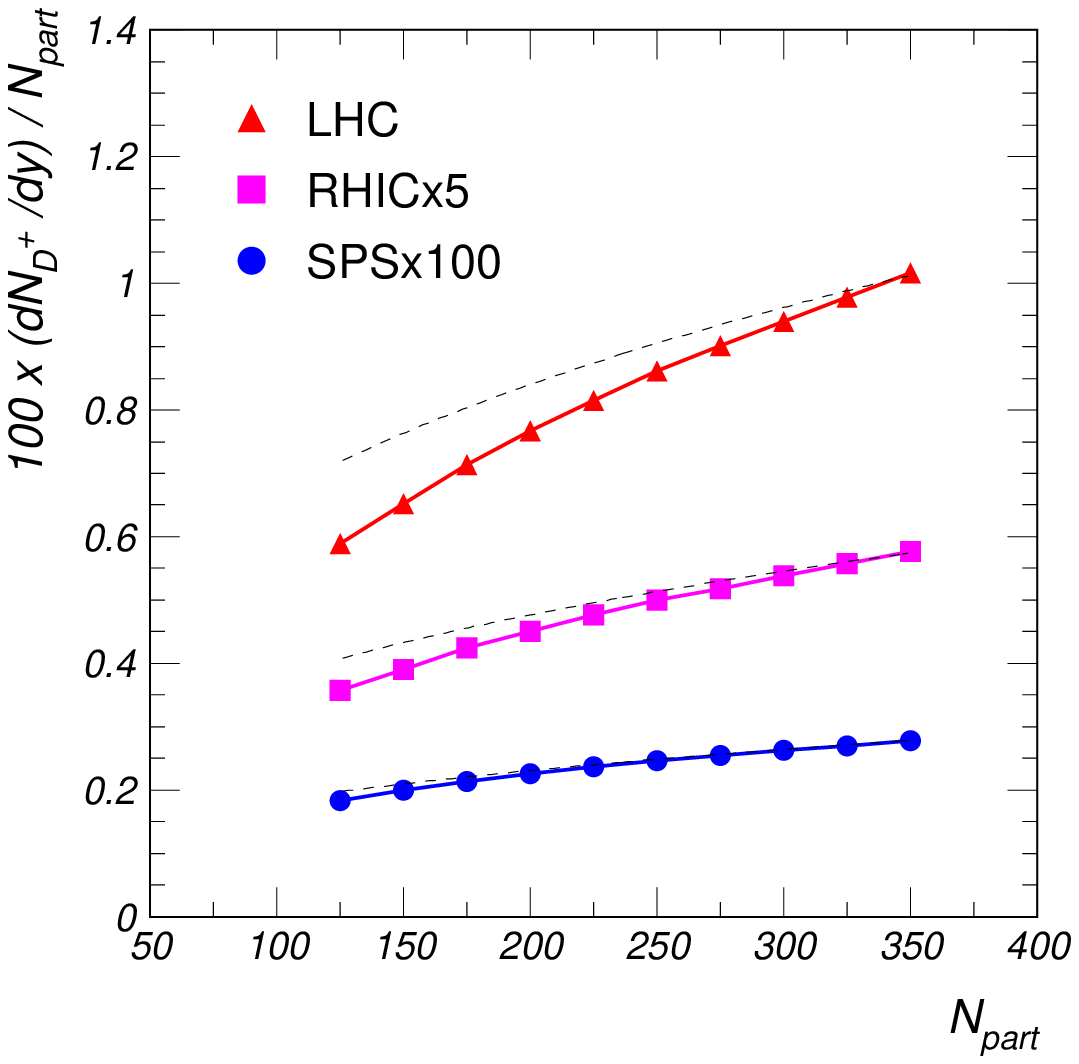}
\vspace{-10mm}
\caption{Centrality dependence of D$^+$ rapidity density at SPS, RHIC 
and LHC.} \label{aa:fig3}
\end{minipage} \end{tabular}
\end{figure}

Predictions for the centrality dependence of $J/\psi$ production are
presented in Fig.~\ref{aa:fig2}. In addition to the dramatic change
in magnitude (note the scale-up by factors of 10 and 100 for RHIC and
SPS energy, respectively) the results exhibit a striking change in
centrality dependence, reflecting the transition from a canonical to a
grand-canonical regime (see \cite{pbm2} for more details). The
preliminary PHENIX results on $J/\psi$ production at RHIC \cite{nagle} agree,
within the still large error bars, with our predictions. A stringent
test of the present model can only be made when high statistics $J/\psi$ data
are available. Another important issue in this respect is the accuracy
of the charm cross section, for which so far only indirect measurements are
available \cite{aver}. In any case, very large suppression factors
as predicted, e.g., by \cite{vogt99} seem not supported by the data.
In Fig.~\ref{aa:fig3} we present the predicted centrality dependence
of charged D$^+$-meson production for the three energies. The expected 
approximate scaling of the ratio D$^+$/$N_{part}$ like $N_{part}^{1/3}$
(dashed lines in Fig.~\ref{aa:fig3}) is only roughly fulfilled due to
departures of the nuclear overlap function from the simple $N_{part}^{4/3}$ 
dependence.

\section{Conclusions}
We have demonstrated that the statistical coalescence approach yields
a good description of the measured centrality dependence of $J/\psi$
production at SPS energy, albeit with a charm cross section increased
by a factor of 2.8 compared to current NLO calculation. Rapidity
densities for open and hidden charm mesons are predicted to increase
strongly with energy, with striking changes in centrality
dependence. First RHIC data on $J/\psi$ production support the current
predictions, although the errors are too large to make firm
conclusions. The statistical coalescence implies travel of charm quarks
over significant distances in QGP. If the model predictions will describe
consistently precision data this would be a clear signal for the presence 
of a deconfined phase.

\end{document}